\documentclass[aps]{revtex4}

\usepackage{graphics}
\usepackage{epsfig}
\usepackage{bm}
\usepackage{amsmath,amssymb}
\usepackage{wasysym}

\def\vb#1{\mbox{\boldmath$#1$}}
\def\pd#1#2{\frac{\partial #1}{\partial #2}}
\def\fd#1#2{\frac{\delta #1}{\delta #2}}
\def\wh#1{\widehat{#1}}
\def\bdot{\,\vb{\cdot}\,}
\def\btimes{\,\vb{\times}\,}

\def\bhat{\wh{{\sf b}}}
\def\cal#1{\mathcal{#1}}

\def\bhat{\wh{{\sf b}}}

\newcommand{\bc}{\begin{center}}
\newcommand{\ec}{\end{center}}
\newcommand{\bt}{\begin{tabbing}}
\newcommand{\et}{\end{tabbing}}
\newcommand{\be}{\begin{equation}}
\newcommand{\ee}{\end{equation}}
\newcommand{\ba}{\begin{eqnarray}}
\newcommand{\ea}{\end{eqnarray}}

\begin{document}

\title{Proof of the Jacobi Property for the guiding-center Vlasov-Maxwell Bracket}

\author{Alain J.~Brizard}
\affiliation{Department of Physics, Saint Michael's College, Colchester, VT 05439, USA}

\begin{abstract}
The proof of the Jacobi property of the guiding-center Vlasov-Maxwell bracket underlying the Hamiltonian structure of the guiding-center Vlasov-Maxwell equations is presented.
\end{abstract}

\date{\today}


\maketitle

\section{Introduction}

The variational formulations of the guiding-center Vlasov-Maxwell equations were presented by Brizard and Tronci \cite{Brizard_Tronci_2016}, whose work also included a derivation of exact conservation laws for energy-momentum and angular momentum through the Noether method. In a companion paper \cite{Brizard_2021}, the guiding-center Vlasov-Maxwell equations are expressed in Hamiltonian form
\begin{eqnarray}
\pd{F_{\rm gc}}{t} &=& -\;\nabla\bdot\left( F_{\rm gc}\;\frac{d{\bf X}}{dt} \right) \;-\; \pd{}{p_{\|}} \left( F_{\rm gc}\;\frac{dp_{\|}}{dt} \right) \;\equiv\; -\;\pd{}{Z^{\alpha}}\left(F_{\rm gc}\;\frac{dZ^{\alpha}}{dt}\right),
\label{eq:V_bracket} \\
 \pd{\bf E}{t} & = & 4\pi c\,\nabla\btimes\fd{{\cal H}_{\rm gc}}{\bf B} - 4\pi q \int_{P}F_{\rm gc}\;\frac{d{\bf X}}{dt} \;-\; \nabla\btimes\left( 4\pi q \int_{P} F_{\rm gc}\;\mathbb{P}_{\|}\bdot\frac{d{\bf X}}{dt} \right),
  \label{eq:B_bracket} \\
\pd{\bf B}{t} &=& -\,4\pi c\;\nabla\btimes\fd{{\cal H}_{\rm gc}}{\bf E},
  \label{eq:E_bracket}
  \end{eqnarray}
where
 \begin{equation}
 \frac{d Z^{\alpha}}{dt} \;=\;  \left\{ Z^{\alpha},\; \fd{{\cal H}_{\rm gc}}{F_{\rm gc}} \right\}_{\rm gc} + 4\pi q \left(\fd{{\cal H}_{\rm gc}}{\bf E} + \mathbb{P}_{\|}\bdot\nabla\btimes\fd{{\cal H}_{\rm gc}}{\bf E}\right)\bdot\{{\bf X},\; Z^{\alpha}\}_{\rm gc}.
 \end{equation}
Here, $q$ denotes the charge of a guiding-center particle, with guiding-center position ${\bf X}$ and guiding-center parallel (kinetic) momentum $p_{\|}$, and summation over charge species is implied wherever an integral of the guiding-center 
phase-space density $F_{\rm gc}$ appears (with $\int_{P}$ denoting an integral over $p_{\|}$ and the magnetic moment $\mu$). In addition, we introduced the symmetric dyadic tensor
\begin{equation}
 \mathbb{P}_{\|} \;\equiv\; \frac{cp_{\|}}{qB}\;\left({\bf I} \;-\; \bhat\,\bhat\right),
 \label{eq:P_par}
 \end{equation}
and the guiding-center Poisson bracket is
\begin{eqnarray}
\{ f,\; g\}_{\rm gc} &\equiv& \frac{{\bf B}^{*}}{B_{\|}^{*}}\bdot\left(\nabla f\;\pd{g}{p_{\|}} \;-\; \pd{f}{p_{\|}}\;\nabla g\right) \;-\; \frac{c\bhat}{qB_{\|}^{*}}\bdot\nabla f\btimes\nabla g \nonumber \\
 &\equiv& \frac{1}{B_{\|}^{*}}\nabla\bdot\left( B_{\|}^{*}\,f\frac{}{}
\{{\bf X},\; g\}_{\rm gc} \right) \;+\; \frac{1}{B_{\|}^{*}}\;\pd{}{p_{\|}}\left( B_{\|}^{*}\,f\frac{}{}
\{p_{\|},\; g\}_{\rm gc} \right),
\label{eq:PB_gc}
\end{eqnarray}
which can also be expressed in divergence form, where we have omitted the ignorable gyromotion pair $(\mu,\theta)$, and
\begin{equation}
\left. \begin{array}{rcl}
{\bf B}^{*} & \equiv & {\bf B} \;+\; (p_{\|}c/q)\,\nabla\btimes\bhat \\
B_{\|}^{*} & \equiv & \bhat\bdot{\bf B}^{*} \;=\; B \;+\; (p_{\|}c/q)\,\bhat\bdot\nabla\btimes\bhat
\end{array} \right\}.
\label{eq:B*_def}
\end{equation}
We note that the guiding-center Poisson bracket \eqref{eq:PB_gc} satisfies the Jacobi property
\begin{equation}
\left\{ \{f, g\}_{\rm gc},\frac{}{} h\right\}_{\rm gc} + \left\{ \{g, h\}_{\rm gc},\frac{}{} f\right\}_{\rm gc} + \left\{ \{h, f\}_{\rm gc},\frac{}{} g\right\}_{\rm gc} = 0,
\label{eq:gcPB_Jacobi}
\end{equation}
which holds for arbitrary functions $(f,g,h)$, subject to the condition
\begin{equation}
\nabla\bdot{\bf B}^{*} \;=\; \nabla\bdot{\bf B} \;=\; 0,
\label{eq:div_Bstar}
\end{equation}
which is satisfied by the definition \eqref{eq:B*_def}. 

Lastly, the guiding-center Hamiltonian functional in Eqs.~\eqref{eq:V_bracket}-\eqref{eq:E_bracket} is
\begin{equation}
{\cal H}_{\rm gc} \;\equiv\; \int_{Z} F_{\rm gc}\;K_{\rm gc} \;+\; \int_{X} \frac{1}{8\pi}\left(|{\bf E}|^{2} + |{\bf B}|^{2}\right),
\label{eq:gc_H}
\end{equation}
so that the Hamiltonian functional derivatives in the guiding-center Vlasov-Maxwell equations \eqref{eq:V_bracket}-\eqref{eq:E_bracket} are
\begin{equation}
\left( \begin{array}{c}
\delta{\cal H}_{\rm gc}/\delta F_{\rm gc} \\
\delta{\cal H}_{\rm gc}/\delta{\bf E} \\
\delta{\cal H}_{\rm gc}/\delta{\bf B}
\end{array} \right) \;=\; \left( \begin{array}{c}
K_{\rm gc} \\
{\bf E}/4\pi \\
{\bf B}/4\pi + \int_{P}F_{\rm gc}\;\mu\,\bhat
\end{array} \right),
\label{eq:delta_H}
\end{equation}
where the guiding-center kinetic energy $K_{\rm gc} = p_{\|}^{2}/2m + \mu\,B$ yields the functional derivative $\delta K_{\rm gc}/\delta{\bf B} = \mu\,\bhat$. 

\section{\label{sec:Jacobi}Guiding-center Vlasov-Maxwell bracket}

The guiding-center Vlasov-Maxwell bracket is initially constructed from the guiding-center Vlasov-Maxwell equations \eqref{eq:V_bracket}-\eqref{eq:E_bracket} and the Hamiltonian functional \eqref{eq:gc_H} as the functional identity
\begin{equation}
\pd{\cal F}{t} \;=\; \left[{\cal F},\frac{}{} {\cal H}_{\rm gc}\right]_{\rm gc} \;=\; \int_{Z}\pd{F_{\rm gc}}{t}\;\fd{\cal F}{F_{\rm gc}} + \int_{X}\left(\pd{\bf E}{t}\bdot\fd{\cal F}{\bf E} + \pd{\bf B}{t}\bdot\fd{\cal F}{\bf B}\right) \;\equiv\; \left\langle \fd{\cal F}{\Psi^{a}}
\left|\frac{}{}\right. \pd{\Psi^{a}}{t}\right\rangle,
\label{eq:F_fd}
\end{equation} 
where the guiding-center Vlasov-Maxwell bracket for two arbitrary functionals $({\cal F},{\cal G})$ of the fields $\vb{\Psi} = (F_{\rm gc},{\bf E},{\bf B})$ is defined in terms of the Poisson structure:
\begin{equation}
\left[{\cal F},\frac{}{} {\cal G}\right]_{\rm gc} \;\equiv\; \left\langle \fd{\cal F}{\Psi^{a}}\;\left|\frac{}{}\right. {\sf J}_{\rm gc}^{ab}(\vb{\Psi})\,\fd{\cal G}{\Psi^{b}} \right\rangle.
\label{eq:Poisson_structure}
\end{equation}
Here, the antisymmetric Poisson operator ${\sf J}_{\rm gc}^{ab}(\vb{\Psi})$ guarantees the antisymmetry property: $[{\cal F},{\cal G}]_{\rm gc} = -\,[{\cal G},{\cal F}]_{\rm gc}$; and the bilinearity of Eq.~\eqref{eq:Poisson_structure} guarantees the Leibnitz property: $[{\cal F},{\cal G}\,{\cal K}]_{\rm gc} = [{\cal F},{\cal G}]_{\rm gc}\,{\cal K} + {\cal G}\,[{\cal F},{\cal K}]_{\rm gc}$. The Jacobi property:
\begin{equation}
{\cal Jac}[{\cal F},{\cal G},{\cal K}] \;\equiv\; \left[[{\cal F},{\cal G}]_{\rm gc},\frac{}{} {\cal K}\right]_{\rm gc} + \left[[{\cal G},{\cal K}]_{\rm gc},\frac{}{} {\cal F}\right]_{\rm gc} + \left[[{\cal K},{\cal F}]_{\rm gc},\frac{}{} {\cal G}\right]_{\rm gc} = 0,
\label{eq:Jacobi}
\end{equation}
which holds for arbitrary functionals $({\cal F},{\cal G},{\cal K})$, involves constraints on the Poisson operator ${\sf J}_{\rm gc}^{ab}(\vb{\Psi})$. The purpose of the present notes is to provide an explicit proof that the guiding-center Vlasov-Maxwell bracket defined in Eq.~\eqref{eq:F_fd} satisfies the Jacobi property \eqref{eq:Jacobi} exactly.

From Eq.~\eqref{eq:F_fd}, we can now extract the guiding-center Vlasov-Maxwell bracket expressed in terms of two arbitrary guiding-center functionals $({\cal F},{\cal G})$ as
 \begin{eqnarray}
 \left[{\cal F},\frac{}{}{\cal G}\right]_{\rm gc} &=& \int_{Z} F_{\rm gc} \left( \left\{ \fd{{\cal F}}{F_{\rm gc}} ,\; \fd{\cal G}{F_{\rm gc}} \right\}_{\rm gc} \;+\; 4\pi q \frac{\delta^{\star}{\cal F}}{\delta{\bf E}}\bdot\left\{{\bf X},\; {\bf X}\right\}_{\rm gc}\bdot 4\pi q \frac{\delta^{\star}{\cal G}}{\delta{\bf E}}\right) \nonumber \\
  &&+\; 4\pi q \int_{Z} F_{\rm gc}\left(  \frac{\delta^{\star}{\cal G}}{\delta{\bf E}}\bdot\left\{{\bf X},\; \fd{{\cal F}}{F_{\rm gc}} \right\}_{\rm gc} -   \frac{\delta^{\star}{\cal F}}{\delta{\bf E}}\bdot\left\{{\bf X},\;\fd{\cal G}{F_{\rm gc}} \right\}_{\rm gc} \right) \nonumber \\
  &&+\; 4\pi c \int_{X} \left(\fd{\cal F}{\bf E}\bdot\nabla\btimes\fd{\cal G}{\bf B} - \fd{\cal G}{\bf E}\bdot\nabla\btimes\fd{\cal F}{\bf B} \right),
 \label{eq:gcVM_bracket}
 \end{eqnarray}
where we introduced the definition
\begin{equation}
\frac{\delta^{\star}{\cal F}}{\delta{\bf E}} \;\equiv\; \fd{\cal F}{\bf E} \;+\; \mathbb{P}_{\|}\bdot\nabla\btimes\fd{\cal F}{\bf E},
\label{eq:fdF_star}
\end{equation}
and
\begin{eqnarray}
\left\{ \fd{{\cal F}}{F_{\rm gc}} ,\; \fd{\cal G}{F_{\rm gc}} \right\}_{\rm gc} &=& \frac{{\bf B}^{*}}{B_{\|}^{*}}\bdot\left(\nabla f\;\pd{g}{p_{\|}} \;-\; \pd{f}{p_{\|}}\;\nabla g\right) \;-\; \frac{c\bhat}{qB_{\|}^{*}}\bdot\left(\nabla f\btimes\nabla g\right), \\
  \frac{\delta^{\star}{\cal G}}{\delta{\bf E}}\bdot\left\{{\bf X},\; \fd{{\cal F}}{F_{\rm gc}} \right\}_{\rm gc} &=& {\bf G}^{\star}\bdot\left(\frac{{\bf B}^{*}}{B_{\|}^{*}}\;\pd{f}{p_{\|}} \;+\; \frac{c\bhat}{qB_{\|}^{*}}\btimes\nabla f \right), \\
 \frac{\delta^{\star}{\cal F}}{\delta{\bf E}}\bdot\left\{{\bf X},\; {\bf X}\right\}_{\rm gc}\bdot  \frac{\delta^{\star}{\cal G}}{\delta{\bf E}} &=& -\,\frac{c\bhat}{qB_{\|}^{*}}\bdot{\bf F}^{\star}\btimes{\bf G}^{\star},
\end{eqnarray}
with the notation $(f,g,k) \equiv (\delta{\cal F}/\delta F_{\rm gc}, \delta{\cal G}/\delta F_{\rm gc}, \delta{\cal K}/\delta F_{\rm gc})$ and $({\bf F}^{\star},{\bf G}^{\star},{\bf K}^{\star}) \equiv (\delta^{\star}{\cal F}/\delta {\bf E}, \delta^{\star}{\cal G}/\delta{\bf E}, \delta^{\star}{\cal K}/\delta  {\bf E})$. It is clear that the bracket \eqref{eq:gcVM_bracket} is antisymmetric and, since it is bilinear in functional derivatives, it satisfies the Leibnitz property.

We now need to verify that the guiding-center bracket \eqref{eq:gcVM_bracket} satisfies the Jacobi property \eqref{eq:Jacobi}. According to the Bracket theorem \cite{Morrison_2013},  the proof of the Jacobi property involves only the explicit dependence of the Poisson operator ${\sf J}_{\rm gc}^{ab}(F_{\rm gc}, {\bf B})$, where we note that the dependence on the magnetic field ${\bf B}$ enters through the guiding-center Poisson bracket \eqref{eq:PB_gc}, while the electric field ${\bf E}$ is explicitly absent. Hence, we can therefore write the double-bracket involving three arbitrary guiding-center functionals $({\cal F}, {\cal G}, {\cal K})$:
\begin{eqnarray}
\left[[{\cal F},{\cal G}]_{\rm gc},\frac{}{} {\cal K}\right]_{\rm gc}^{\rm P} &=& \int_{Z} F_{\rm gc} \left\{ \frac{\delta^{P}[{\cal F}, {\cal G}]_{\rm gc}}{\delta F_{\rm gc}}, \fd{\cal K}{F_{\rm gc}} \right\}_{\rm gc} +\, 4\pi q \int_{Z} F_{\rm gc}\; \frac{\delta^{\star}{\cal K}}{\delta{\bf E}}
\bdot\left\{ {\bf X},  \frac{\delta^{P}[{\cal F}, {\cal G}]_{\rm gc}}{\delta F_{\rm gc}} \right\}_{\rm gc} \nonumber \\
 &&-\; 4\pi c \int_{X}  \frac{\delta^{P}[{\cal F}, {\cal G}]_{\rm gc}}{\delta {\bf B}}\bdot\nabla\btimes\fd{\cal K}{\bf E},
 \label{eq:Jac_fg-k}
\end{eqnarray}
where the Poisson functional derivative $\delta^{P}[{\cal F},{\cal G}]/\delta F_{\rm gc}$ only involves variations of the Poisson operator $\partial{\sf J}_{\rm gc}^{ab}(F_{\rm gc}, {\bf B})/\partial F_{\rm gc}$ in the Vlasov and Interaction sub-brackets, while $\delta^{P}[{\cal F}, {\cal G}]_{\rm gc}/\delta{\bf B}$ in the Maxwell sub-bracket involves the explicit dependence of ${\sf J}_{\rm gc}^{ab}(F_{\rm gc}, {\bf B})$ on the magnetic field ${\bf B}$ appearing through $(\bhat,{\bf B}^{*},B_{\|}^{*})$ in the guiding-center Poisson bracket \eqref{eq:PB_gc} and the dyadic tensor \eqref{eq:P_par}, where
\begin{equation}
\left. \begin{array}{rcl}
\delta{\bf B}^{*} &=& \delta{\bf B} + \nabla\btimes(\mathbb{P}_{\|}\bdot\delta{\bf B}) \\
(c/q)\,\delta\bhat &=& \delta{\bf B}\bdot\partial\mathbb{P}_{\|}/\partial p_{\|}
\end{array} \right\},
\label{eq:deltaB_bhat}
\end{equation}
with $\mathbb{P}_{\|}$ defined in Eq.~\eqref{eq:P_par}.

\subsection{Vlasov and Interaction sub-brackets}

From the guiding-center Vlasov-Maxwell bracket \eqref{eq:gcVM_bracket}, we find the Poisson functional derivative
\begin{eqnarray}
 \frac{\delta^{P}[{\cal F}, {\cal G}]_{\rm gc}}{\delta F_{\rm gc}} &=& \left\{ f,\; g \right\}_{\rm gc} \;-\; 4\pi q \left({\bf F}^{\star}\bdot\left\{{\bf X},\;g\right\}_{\rm gc} -  {\bf G}^{\star}\bdot\left\{{\bf X},\;f\right\}_{\rm gc} \right) \;+\; (4\pi q)^{2} \left(  {\bf F}^{\star}
 \bdot\left\{{\bf X},\; {\bf X}\right\}_{\rm gc}\bdot  {\bf G}^{\star}\right),
  \label{eq:gcVM_double}
\end{eqnarray}
where the functional derivatives $(f,g; {\bf F}^{\star},{\bf G}^{\star})$ are left intact according to the Bracket Theorem. Hence, the Vlasov sub-bracket in Eq.~\eqref{eq:Jac_fg-k} includes the terms
\begin{eqnarray}
 \left\{ \frac{\delta^{P}[{\cal F}, {\cal G}]_{\rm gc}}{\delta F_{\rm gc}}, \fd{\cal K}{F_{\rm gc}} \right\}_{\rm gc} &=&  \left\{  \left\{ f,\;  g \right\}_{\rm gc},  k \right\}_{\rm gc} \;-\; 4\pi q  \left\{ \left( {\bf F}^{\star}\bdot\left\{{\bf X},\;  g \right\}_{\rm gc} -  {\bf G}^{\star}\bdot\left\{
 {\bf X},\; f \right\}_{\rm gc} \right),  k \right\}_{\rm gc} \nonumber \\
  &&+\; (4\pi q)^{2} \left\{ \left(  {\bf F}^{\star}\bdot\left\{{\bf X},\; {\bf X}\right\}_{\rm gc}\bdot  {\bf G}^{\star}\right),  k \right\}_{\rm gc},
  \label{eq:VInt_1}
 \end{eqnarray}
 while the Interaction sub-bracket  in Eq.~\eqref{eq:Jac_fg-k} includes the terms
 \begin{eqnarray}
4\pi q\, \frac{\delta^{\star}{\cal K}}{\delta{\bf E}}\bdot \left\{ {\bf X},\; \frac{\delta^{P}[{\cal F}, {\cal G}]_{\rm gc}}{\delta F_{\rm gc}} \right\}_{\rm gc} &=&  4\pi q\, {\bf K}^{\star}\bdot \left\{ {\bf X},\; \left\{ f,\;  g\right\}_{\rm gc}\right\}_{\rm gc} \;-\; (4\pi q)^{2}  {\bf K}^{\star}\bdot 
\left\{ {\bf X},\; \left( {\bf F}^{\star}\bdot\left\{{\bf X},\;  g \right\}_{\rm gc} -   {\bf G}^{\star}\bdot\left\{{\bf X},\; f \right\}_{\rm gc} \right)\right\}_{\rm gc} \nonumber \\
 &&+\; (4\pi q)^{3}  {\bf K}^{\star}\bdot \left\{ {\bf X},\; \left( {\bf F}^{\star}\bdot\left\{{\bf X},\; {\bf X}\right\}_{\rm gc}\bdot {\bf G}^{\star}\right) \right\}_{\rm gc}.
 \label{eq:VInt_2}
 \end{eqnarray}
 We note, here, that the ordering $(4\pi q)^{n}$, with $0 \leq n \leq 3$, will be useful in verifying the Jacobi property of the guiding-center Vlasov-Maxwell bracket \eqref{eq:gcVM_bracket}, i.e., the Jacobi property must hold separately for each power $n$.
 
 At the zeroth order in $4\pi q$ in Eqs.~\eqref{eq:VInt_1}-\eqref{eq:VInt_2}, the Vlasov-Maxwell contributions to the guiding-center Jacobi property \eqref{eq:Jacobi} are
 \begin{equation}
\int_{Z} F_{\rm gc} \left( \left\{ \{ f,\; g\}_{\rm gc},\frac{}{} k \right\}_{\rm gc} + \left\{ \{ g,\; k\}_{\rm gc},\frac{}{} f \right\}_{\rm gc} + \left\{ \{ k,\; f\}_{\rm gc},\frac{}{} g \right\}_{\rm gc} \right) \;\equiv\; \int_{Z} F_{\rm gc}\,{\sf Jac}_{VM}^{(0)}[f,g,k] .
\label{eq:Vlasov_0}
\end{equation}
Next, at the first order in $4\pi q$ in Eqs.~\eqref{eq:VInt_1}-\eqref{eq:VInt_2}, the Vlasov-Interaction (VI) contributions to the guiding-center Jacobi property \eqref{eq:Jacobi} are
\begin{equation}
4\pi q \int_{Z} F_{\rm gc} \left( {\sf Jac}_{VI}^{(1)}[f,g;{\bf K}^{\star}] \;+\; {\sf Jac}_{VI}^{(1)}[g,k;{\bf F}^{\star}] \;+\; {\sf Jac}_{VI}^{(1)}[k,f;{\bf G}^{\star}]  \right),
\label{eq:VI_1}
\end{equation}
where
\begin{eqnarray}
{\sf Jac}_{VI}^{(1)}[f,g;{\bf K}^{\star}] &=& {\bf K}^{\star}\bdot\left\{ {\bf X},\frac{}{} \{ f, g\}_{\rm gc}\right\}_{\rm gc} \;+\; \left\{ {\bf K}^{\star}\bdot\{{\bf X}, g\}_{\rm gc},\frac{}{} f \right\}_{\rm gc} \;-\; \left\{ {\bf K}^{\star}\bdot\{{\bf X}, f\}_{\rm gc},\frac{}{} g 
\right\}_{\rm gc} \nonumber \\
 &=& {\sf Jac}_{VM}^{(1)}[f,g;{\bf K}^{*}] \;+\;  \left\{ {\bf K}^{\star},\; f\right\}_{\rm gc}\bdot\{{\bf X},\; g \}_{\rm gc} - \left\{ {\bf K}^{\star},\; g\right\}_{\rm gc}\bdot\{{\bf X},\; f \}_{\rm gc},
  \label{eq:VI_1_final}
\end{eqnarray}
which is obtained after using the Leibnitz formula: 
\[ \left\{{\bf K}^{\star}\bdot\{{\bf X}, g\}_{\rm gc},\frac{}{} f\right\}_{\rm gc} \;=\; \{ {\bf K}^{\star}, f\}_{\rm gc}\bdot\{{\bf X}, g\}_{\rm gc} \;+\; {\bf K}^{\star}\bdot\left\{\{{\bf X},g\}_{\rm gc},\frac{}{} f\right\}_{\rm gc}, \]
and the first-order Vlasov-Maxwell (VM) contribution is defined as
\begin{equation}
{\sf Jac}_{VM}^{(1)}[f,g;{\bf K}^{*}] \;\equiv\; K_{i}^{\star}\bdot\left( \left\{ X^{i},\frac{}{} \{ f, g\}_{\rm gc}\right\}_{\rm gc} + \left\{ f,\frac{}{} \{g,\; X^{i}\}_{\rm gc} \right\}_{\rm gc} +  \left\{ g,\frac{}{} \{X^{i},f \}_{\rm gc} \right\}_{\rm gc}\right). 
\label{eq:MV_1}
\end{equation}
At the second order in $4\pi q$ in Eqs.~\eqref{eq:VInt_1}-\eqref{eq:VInt_2}, the Vlasov-Interaction contributions to the guiding-center Jacobi property \eqref{eq:Jacobi} are
\begin{equation}
(4\pi q)^{2} \int_{Z} F_{\rm gc} \left( {\sf Jac}_{VI}^{(2)}[f;{\bf G}^{\star},{\bf K}^{\star}] \;+\; {\sf Jac}_{VI}^{(2)}[g;{\bf K}^{\star},{\bf F}^{\star}] \;+\; {\sf Jac}_{VI}^{(2)}[k;{\bf F}^{\star},{\bf G}^{\star}]  \right),
\label{eq:VI_2}
\end{equation}
where
\begin{eqnarray}
{\sf Jac}_{VI}^{(2)}[f;{\bf G}^{\star},{\bf K}^{\star}] &=& {\bf K}^{\star}\bdot\left\{ {\bf X},\frac{}{} {\bf G}^{\star}\bdot\{{\bf X},\; f\}_{\rm gc}\right\}_{\rm gc} - {\bf G}^{\star}\bdot\left\{ {\bf X},\frac{}{} {\bf K}^{\star}\bdot\{{\bf X},\; f\}_{\rm gc}\right\}_{\rm gc} 
+ \left\{ {\bf G}^{\star}\bdot\{{\bf X},{\bf X}\}_{\rm gc}\bdot{\bf K}^{\star},\frac{}{} f \right\}_{\rm gc} \nonumber \\
 &=& {\sf Jac}_{VM}^{(2)}[f;{\bf G}^{\star},{\bf K}^{\star}] \;+\; \left\{ {\bf G}^{\star},\; f\right\}_{\rm gc}\bdot\{{\bf X},{\bf X}\}_{\rm gc}\bdot{\bf K}^{\star} \;+\; {\bf G}^{\star}\bdot\{{\bf X},{\bf X}\}_{\rm gc}\bdot\left\{{\bf K}^{\star},\; f\right\}_{\rm gc} \nonumber \\
   &&+\; \left({\bf K}^{\star}\bdot\left\{{\bf X}, {\bf G}^{\star}\right\}_{\rm gc} - {\bf G}^{\star}\bdot\left\{{\bf X}, {\bf K}^{\star}\right\}_{\rm gc} \right)\bdot\{{\bf X},\; f\}_{\rm gc},
   \label{eq:VI_2_final}
\end{eqnarray}
with the second-order Vlasov-Maxwell contribution defined as
\begin{equation}
{\sf Jac}_{VM}^{(2)}[f;{\bf G}^{\star},{\bf K}^{\star}] \;\equiv\; G_{i}^{*} K_{j}^{*} \left( \left\{ X^{j},\frac{}{} \{ X^{i},\; f\}_{\rm gc}\right\}_{\rm gc} + \left\{ X^{i},\frac{}{} \{ f,\; X^{j}\}_{\rm gc}\right\}_{\rm gc} + \left\{ f,\frac{}{} \{ X^{j},\; X^{i}\}_{\rm gc}\right\}_{\rm gc} 
\right).
\label{eq:MV_2}
\end{equation}
Lastly, at the third order in $4\pi q$ in Eqs.~\eqref{eq:VInt_1}-\eqref{eq:VInt_2}, the Vlasov-Interaction contributions to the guiding-center Jacobi property \eqref{eq:Jacobi} are
\begin{equation}
(4\pi q)^{3} \int_{Z} F_{\rm gc}\; {\sf Jac}_{VI}^{(3)}[{\bf F}^{\star},{\bf G}^{\star},{\bf K}^{\star}],
\label{eq:VI_3}
\end{equation}
where
\begin{eqnarray}
{\sf Jac}_{VI}^{(3)}[{\bf F}^{\star},{\bf G}^{\star},{\bf K}^{\star}] &=& {\bf F}^{\star}\bdot\left\{ {\bf X},\frac{}{} {\bf G}^{\star}\bdot\{{\bf X},{\bf X}\}_{\rm gc}\bdot{\bf K}^{\star} \right\}_{\rm gc} \;+\; {\bf G}^{\star}\bdot\left\{ {\bf X},\frac{}{} {\bf K}^{\star}\bdot\{{\bf X},
{\bf X}\}_{\rm gc}\bdot{\bf F}^{\star} \right\}_{\rm gc} \nonumber \\
 &&+\; {\bf K}^{\star}\bdot\left\{ {\bf X},\frac{}{} {\bf F}^{\star}\bdot\{{\bf X},{\bf X}\}_{\rm gc}\bdot{\bf G}^{\star} \right\}_{\rm gc} \nonumber \\
 &=& {\sf Jac}_{VM}^{(3)}[{\bf F}^{\star},{\bf G}^{\star},{\bf K}^{\star}] \;+\; \left({\bf F}^{\star}\bdot\left\{{\bf X}, {\bf G}^{\star}\right\}_{\rm gc} - {\bf G}^{\star}\bdot\left\{{\bf X}, {\bf F}^{\star}\right\}_{\rm gc} \right)\bdot\{{\bf X}, {\bf X}\}_{\rm gc}\bdot{\bf K}^{\star} \nonumber \\
  &&+\; \left({\bf G}^{\star}\bdot\left\{{\bf X}, {\bf K}^{\star}\right\}_{\rm gc} - {\bf K}^{\star}\bdot\left\{{\bf X}, {\bf G}^{\star}\right\}_{\rm gc} \right)\bdot\{{\bf X}, {\bf X}\}_{\rm gc}\bdot{\bf F}^{\star} \nonumber \\
  &&+\; \left({\bf K}^{\star}\bdot\left\{{\bf X}, {\bf F}^{\star}\right\}_{\rm gc} - {\bf F}^{\star}\bdot\left\{{\bf X}, {\bf K}^{\star}\right\}_{\rm gc} \right)\bdot\{{\bf X}, {\bf X}\}_{\rm gc}\bdot{\bf G}^{\star},
 \label{eq:VI_3_final}
\end{eqnarray}
with the third-order Vlasov-Maxwell contribution defined as
\begin{equation}
 {\sf Jac}_{VM}^{(3)}[{\bf F}^{\star},{\bf G}^{\star},{\bf K}^{\star}] \;\equiv\; F_{i}^{*}G_{i}^{*} K_{\ell}^{*} \left( \left\{ X^{i},\frac{}{} \{ X^{j},\; X^{\ell}\}_{\rm gc}\right\}_{\rm gc} + \left\{ X^{j},\frac{}{} \{ X^{\ell},\; X^{i}\}_{\rm gc}\right\}_{\rm gc} + \left\{ X^{\ell},\frac{}{} \{ X^{i},\; X^{j}\}_{\rm gc}\right\}_{\rm gc} \right).
 \label{eq:MV_3}
 \end{equation}

\subsection{Jacobian term}

The next set of terms in Eq.~\eqref{eq:Jac_fg-k} come from contributions due to the magnetic variations of the guiding-center Jacobian, where $\delta B_{\|}^{*}$ is given by Eq.~\eqref{eq:deltaB_star}. The Jacobian contributions to the Jacobi property \eqref{eq:Jacobi} are given by Eq.~\eqref{eq:Maxwell_Jac}:
 \begin{equation}
4\pi q \int_{Z} F_{\rm gc} \left( {\sf Jac}_{J}^{(1)}[f,g; {\bf K}^{\star}] + \leftturn \right) + (4\pi q)^{2} \int_{Z} F_{\rm gc} \left( {\sf Jac}_{J}^{(2)}[f;{\bf G}^{\star}, {\bf K}^{\star}] + \leftturn \right) + (4\pi q)^{3} \int_{Z} F_{\rm gc}\;
 {\sf Jac}_{J}^{(3)}[{\bf F}^{\star},{\bf G}^{\star},{\bf K}^{\star}], 
\end{equation}
where $\leftturn$ denotes cyclic permutations of the functionals $({\cal F},{\cal G},{\cal K})$ and the Jacobian contributions are
\begin{eqnarray}
{\sf Jac}_{J}^{(1)}[f,g; {\bf K}^{\star}] &=& \left\{ X^{i}, K_{i}^{\star}\right\}_{\rm gc}\; \{ f, g\}_{\rm gc}, \label{eq:Jac_1} \\
 {\sf Jac}_{J}^{(2)}[f;{\bf G}^{\star}, {\bf K}^{\star}] &=&  \left\{ X^{i}, K_{i}^{\star}\right\}_{\rm gc}\;{\bf G}^{\star}\bdot\{{\bf X},\;f\}_{\rm gc} \;-\;  \left\{ X^{i}, G_{i}^{*}\right\}_{\rm gc}\;{\bf K}^{\star}\bdot\{{\bf X},\;f\}_{\rm gc}, \label{eq:Jac_2} \\
 {\sf Jac}_{J}^{(3)}[{\bf F}^{\star},{\bf G}^{\star},{\bf K}^{\star}] &=&  \left\{ X^{i}, K_{i}^{\star}\right\}_{\rm gc}{\bf F}^{\star}\bdot\{{\bf X},{\bf X}\}_{\rm gc}\bdot{\bf G}^{\star} \;+\; \left\{ X^{i}, F_{i}^{\star}\right\}_{\rm gc}{\bf G}^{\star}\bdot\{{\bf X},{\bf X}\}_{\rm gc}\bdot{\bf K}^{\star} \nonumber \\
  &&+\; \left\{ X^{i}, G_{i}^{\star}\right\}_{\rm gc}{\bf K}^{\star}\bdot\{{\bf X},{\bf X}\}_{\rm gc}\bdot{\bf F}^{\star}, \label{eq:Jac_3}
 \end{eqnarray} 
 with $\{X^{i},\,K_{i}^{\star}\}_{\rm gc}$ defined in Eq.~\eqref{eq:Id_gc_Jac}.

\subsection{Maxwell sub-bracket}

The calculation of the Poisson functional derivative $\delta^{P}[{\cal F}, {\cal G}]_{\rm gc}/\delta{\bf B}$ in Eq.~\eqref{eq:Jac_fg-k} requires an extensive series of steps.  Here, we derive the Poisson functional derivative of the guiding-center Vlasov-Maxwell bracket \eqref{eq:gcVM_bracket}:
 \begin{eqnarray}
\delta^{P} \left[{\cal F},\frac{}{}{\cal G}\right]_{\rm gc} &=&  \int_{Z} \frac{F_{\rm gc}}{B_{\|}^{*}}\;\delta^{P}\left( B_{\|}^{*}\;\left\{ \fd{{\cal F}}{F_{\rm gc}} ,\; \fd{\cal G}{F_{\rm gc}} \right\}_{\rm gc}\right) \nonumber \\
   &&+\; 4\pi q \int_{Z} \frac{F_{\rm gc}}{B_{\|}^{*}}\;\delta^{P}\left[ B_{\|}^{*}\;\left(  \frac{\delta^{\star}{\cal G}}{\delta{\bf E}}\bdot\left\{{\bf X},\; \fd{{\cal F}}{F_{\rm gc}} \right\}_{\rm gc} -   \frac{\delta^{\star}{\cal F}}{\delta{\bf E}}\bdot\left\{{\bf X},\;\fd{\cal G}{F_{\rm gc}} \right\}_{\rm gc} \right) \right] \nonumber \\
  &&+\; (4\pi q)^{2}  \int_{Z} \frac{F_{\rm gc}}{B_{\|}^{*}}\;\delta^{P}\left( B_{\|}^{*}\;  \frac{\delta^{\star}{\cal F}}{\delta{\bf E}}\bdot\left\{{\bf X},\; {\bf X}\right\}_{\rm gc}\bdot  \frac{\delta^{\star}{\cal G}}{\delta{\bf E}}\right),
  \label{eq:gcVM_bracket_app}
 \end{eqnarray}
 where the Jacobian variations \eqref{eq:Jac_1}-\eqref{eq:Jac_3} are excluded in Eq.~\eqref{eq:gcVM_bracket_app}.
 
 \subsubsection{Zeroth-order term}
 
For the zeroth-order term in Eq.~\eqref{eq:gcVM_bracket_app}, we begin with Eq.~\eqref{eq:Maxwell_zero}:
 \begin{equation}
  \int_{X} \nabla\btimes{\bf K}\bdot\frac{\delta^{P}}{\delta{\bf B}}\left(\int_{Z}  F_{\rm gc} \left\{ f,\; g \right\}_{\rm gc}\right) \;=\; \int_{Z} F \left[ \nabla\btimes{\bf K}^{\star}\bdot\left(\nabla f\,\pd{g}{p_{\|}} - \pd{f}{p_{\|}}\,\nabla g\right) - \pd{{\bf K}^{\star}}{p_{\|}}\bdot\nabla f\btimes\nabla g \right].
 \label{eq:Maxwell_1}
\end{equation}
We now use the identity \eqref{eq:Id_gc_curl}, so that we may write
\begin{equation}
F\;\nabla\btimes{\bf K}^{\star}\bdot\left(\nabla f\,\pd{g}{p_{\|}} - \pd{f}{p_{\|}}\,\nabla g\right) \;=\; \frac{q}{c}\,F_{\rm gc}\;\nabla\btimes{\bf K}^{\star}\bdot\left(\left\{{\bf X},\; f\right\}_{\rm gc}\btimes\left\{ {\bf X},\; g \right\}_{\rm gc} \right) \;+\; F\,\bhat\bdot\nabla\btimes
{\bf K}^{\star}\;\{ f,\; g\}_{\rm gc},
\end{equation}
where the first term on the right side can be written as
\begin{eqnarray}
\nabla\btimes{\bf K}^{\star}\bdot\left(\left\{{\bf X},\; f\right\}_{\rm gc}\btimes\left\{ {\bf X},\; g \right\}_{\rm gc} \right) &=& \left\{ {\bf X},\; f\right\}_{\rm gc}\bdot\nabla{\bf K}^{\star}\bdot\left\{ {\bf X},\; g\right\}_{\rm gc} \;-\; \left\{ {\bf X},\; g\right\}_{\rm gc}\bdot\nabla{\bf K}^{\star}\bdot\left\{ {\bf X},\; f\right\}_{\rm gc} \nonumber \\
 &=& \left\{ {\bf K}^{\star},\; f\right\}_{\rm gc}\bdot \left\{ {\bf X},\; g\right\}_{\rm gc}  \;-\; \left\{ {\bf K}^{\star},\; g\right\}_{\rm gc}\bdot \left\{ {\bf X},\; f\right\}_{\rm gc}  \nonumber \\
  &&+\; \pd{{\bf K}^{\star}}{p_{\|}}\bdot\left( \left\{ {\bf X},\; g\right\}_{\rm gc} \;\frac{{\bf B}^{*}}{B_{\|}^{*}}\bdot\nabla f \;-\; \left\{ {\bf X},\; f\right\}_{\rm gc} \;\frac{{\bf B}^{*}}{B_{\|}^{*}}\bdot\nabla g\right).
\end{eqnarray}
Next, we use the identity \eqref{eq:Id_gc_Rf} to write
\begin{equation}
F\,\bhat\bdot\nabla\btimes{\bf K}^{\star} \;=\; \frac{q}{c}\,F_{\rm gc} \left( \left\{ X^{i},\; K_{i}^{\star}\right\}_{\rm gc} \;-\; \frac{{\bf B}^{*}}{B_{\|}^{*}}\bdot\pd{{\bf K}^{\star}}{p_{\|}}\right),
\end{equation}
and, after combining these expressions in Eq.~\eqref{eq:Maxwell_1}, we obtain
\begin{equation}
 -\,4\pi c \int_{X} \nabla\btimes{\bf K}\bdot\frac{\delta^{P}}{\delta{\bf B}}\left(\int_{Z}  F_{\rm gc} \left\{ f,\; g \right\}_{\rm gc}\right)  \;=\; 4\pi q\int_{Z} F_{\rm gc}\;{\sf Jac}_{M}^{(1)}[f,g; {\bf K}^{\star}].
\end{equation}
Here, all terms associated with $\partial {\bf K}^{\star}/\partial p_{\|}$ have canceled out:
\[ \int_{Z} F\,\pd{{\bf K}^{\star}}{p_{\|}}\bdot\left[ \left(\frac{c}{q}\right)\nabla f\btimes\nabla g + {\bf B}^{*}\,\{f,g\}_{\rm gc} + \{{\bf X},f\}_{\rm gc}\,({\bf B}^{*}\bdot\nabla g) - \{{\bf X},g\}_{\rm gc}\,({\bf B}^{*}\bdot\nabla f) \right] \;=\; 0, \] 
and the first-order Maxwell sub-bracket contribution is
\begin{equation}
{\sf Jac}_{M}^{(1)}[f,g; {\bf K}^{\star}] \;=\; -\,\left\{ {\bf K}^{\star},\frac{}{} f\right\}_{\rm gc}\bdot\{{\bf X}, g\}_{\rm gc} \;+\; \left\{ {\bf K}^{\star},\frac{}{} g\right\}_{\rm gc}\bdot\{{\bf X}, f\}_{\rm gc} \;-\; \{f,\; g\}_{\rm gc}\;\left\{ X^{i},\frac{}{} K_{i}^{\star}\right\}_{\rm gc}.
\label{eq:Maxwell_1_final}
\end{equation}
Hence, by combining Eqs.~\eqref{eq:VI_1_final}, \eqref{eq:Jac_1}, and \eqref{eq:Maxwell_1_final}, we finally obtain
\begin{equation}
{\sf Jac}_{VI}^{(1)}[f,g; {\bf K}^{\star}] + {\sf Jac}_{J}^{(1)}[f,g; {\bf K}^{\star}] + {\sf Jac}_{M}^{(1)}[f,g; {\bf K}^{\star}] \;=\; {\sf Jac}_{VM}^{(1)}[f,g; {\bf K}^{\star}],
\label{eq:VM_1}
\end{equation}
where the first-order Vlasov-Maxwell term ${\sf Jac}_{VM}^{(1)}[f,g; {\bf K}^{\star}]$ is defined in Eq.~\eqref{eq:MV_1}.

 \subsubsection{First-order term}
 
For the first-order term in $4\pi q$ in Eq.~\eqref{eq:gcVM_bracket_app}, we  begin with Eq.~\eqref{eq:Maxwell_one}:
 \begin{eqnarray}
 && \int_{X}\nabla\btimes{\bf K}\bdot\frac{\delta^{P}}{\delta{\bf B}}\left[\int_{Z} F_{\rm gc}\,\left({\bf G}^{\star}\bdot\{ {\bf X},\; f\}_{\rm gc} -\frac{}{} {\bf F}^{\star}\bdot\{ {\bf X},\; g\}_{\rm gc} \right) \right] \nonumber \\
  &&\hspace*{0.3in}=\; \int_{Z} F\left[ \nabla\btimes{\bf K}^{\star}\bdot\left( {\bf G}^{\star}\;\pd{f}{p_{\|}} \;-\; {\bf F}^{\star}\;\pd{g}{p_{\|}} \right) \;-\; \pd{{\bf K}^{\star}}{p_{\|}}\bdot\left({\bf G}^{\star}\btimes\nabla f \;-\frac{}{} {\bf F}^{\star}\btimes\nabla g\right)
  \right].
  \label{eq:Maxwell_1_a}
\end{eqnarray}
First, using the identity $\partial f/\partial p_{\|} = \bhat\bdot\{{\bf X}, f\}_{\rm gc}$, we can write
\[ {\bf G}^{\star}\bdot\nabla\btimes{\bf K}^{\star} \left(\bhat\bdot\{{\bf X}, f\}_{\rm gc}\right) \;=\; \left({\bf G}^{\star}\btimes\bhat\right)\bdot\left( \nabla\btimes{\bf K}^{\star}\btimes\{{\bf X},\, f\}_{\rm gc}\right) + \left(\bhat\bdot\nabla\btimes{\bf K}^{\star}\right)\;
{\bf G}^{\star}\bdot\{{\bf X},\, f\}_{\rm gc}, \]
so that
\begin{eqnarray}
\nabla\btimes{\bf K}^{\star}\bdot\left( {\bf G}^{\star}\;\pd{f}{p_{\|}} \;-\; {\bf F}^{\star}\;\pd{g}{p_{\|}} \right) &=& \nabla\btimes{\bf K}^{\star}\bdot\left[ \{{\bf X},\, f\}_{\rm gc}\btimes\left({\bf G}^{\star}\btimes\bhat\right) \;-\; 
\{{\bf X},\, g\}_{\rm gc}\btimes\left({\bf F}^{\star}\btimes\bhat\right)\right] \nonumber \\
 &&+\; \bhat\bdot\nabla\btimes{\bf K}^{\star}\; \left({\bf G}^{\star}\bdot\{{\bf X},\, f\}_{\rm gc} \;-\frac{}{} {\bf F}^{\star}\bdot\{{\bf X},\, g\}_{\rm gc} \right).
\label{eq:Maxwell_1_b}
 \end{eqnarray}
 Here, the last term on the right side of Eq.~\eqref{eq:Maxwell_1_b} can be written as
 \begin{eqnarray}
 \bhat\bdot\nabla\btimes{\bf K}^{\star}\; \left({\bf G}^{\star}\bdot\{{\bf X},\, f\}_{\rm gc} \;-\frac{}{} {\bf F}^{\star}\bdot\{{\bf X},\, g\}_{\rm gc} \right) &=& \frac{qB_{\|}^{*}}{c}\;\left\{X^{i}, K_{i}^{\star}\right\}_{\rm gc}\;\left({\bf G}^{\star}\bdot\{{\bf X},\, f\}_{\rm gc} 
 \;-\frac{}{} {\bf F}^{\star}\bdot\{{\bf X},\, g\}_{\rm gc} \right) \nonumber \\
  &&-\; \frac{q}{c}\,{\bf B}^{*}\bdot \pd{{\bf K}^{\star}}{p_{\|}}\;\left({\bf G}^{\star}\bdot\{{\bf X},\, f\}_{\rm gc} \;-\frac{}{} {\bf F}^{\star}\bdot\{{\bf X},\, g\}_{\rm gc} \right).
 \end{eqnarray}
 Next, we write
 \begin{eqnarray}
\nabla\btimes{\bf K}^{\star}\bdot\left[\{{\bf X},\, f\}_{\rm gc}\btimes\left({\bf G}^{\star}\btimes\bhat\right) \right] &=& \left(\left\{{\bf K}^{\star},\, f\right\}_{\rm gc} \;+\; \frac{{\bf B}^{*}}{B_{\|}^{*}}\bdot\nabla f\; \pd{{\bf K}^{\star}}{p_{\|}}\right)\bdot\left({\bf G}^{\star}\btimes\bhat\right) \;-\; \left({\bf G}^{\star}\btimes\bhat\right)\bdot\nabla{\bf K}^{\star}\bdot\{{\bf X},\, f\}_{\rm gc} \nonumber \\
  &=& -\,\frac{qB_{\|}^{*}}{c}\;\left(\left\{{\bf K}^{\star},\, f\right\}_{\rm gc}\bdot\{{\bf X},{\bf X}\}_{\rm gc}\bdot{\bf G}^{\star} \;+\; {\bf G}^{\star}\bdot\left\{{\bf X},\, {\bf K}^{\star}\right\}_{\rm gc}\bdot\{{\bf X},\, f \}_{\rm gc} \right) \nonumber \\
   &&+\; \pd{{\bf K}^{\star}}{p_{\|}}\bdot\left[  \left({\bf G}^{\star}\btimes\bhat\right)\; \frac{{\bf B}^{*}}{B_{\|}^{*}}\bdot\nabla f \;+\; \left(\frac{q}{c}\,{\bf B}^{*}\bdot{\bf G}^{\star}\right) \{{\bf X},\, f\}_{\rm gc} \right],
\end{eqnarray}
so that Eq.~\eqref{eq:Maxwell_1_b} becomes
 \begin{eqnarray}
F\,\nabla\btimes{\bf K}^{\star}\bdot\left( {\bf G}^{\star}\;\pd{f}{p_{\|}} \;-\; {\bf F}^{\star}\;\pd{g}{p_{\|}} \right) &=& \frac{q}{c}\,F_{\rm gc} \left[ \left\{X^{i}, K_{i}^{\star}\right\}_{\rm gc}\;\left({\bf G}^{\star}\bdot\{{\bf X},\, f\}_{\rm gc} 
 \;-\frac{}{} {\bf F}^{\star}\bdot\{{\bf X},\, g\}_{\rm gc} \right) \right] \nonumber \\
  &&+\; \frac{q}{c}\,F_{\rm gc} \left(\left\{{\bf K}^{\star},\, g\right\}_{\rm gc}\bdot\{{\bf X},{\bf X}\}_{\rm gc}\bdot{\bf F}^{\star} \;+\; {\bf F}^{\star}\bdot\left\{{\bf X},\, {\bf K}^{\star}\right\}_{\rm gc}\bdot\{{\bf X},\, g \}_{\rm gc} \right) \nonumber \\
  &&-\; \frac{q}{c}\,F_{\rm gc}  \left(\left\{{\bf K}^{\star},\, f\right\}_{\rm gc}\bdot\{{\bf X},{\bf X}\}_{\rm gc}\bdot{\bf G}^{\star} \;+\; {\bf G}^{\star}\bdot\left\{{\bf X},\, {\bf K}^{\star}\right\}_{\rm gc}\bdot\{{\bf X},\, f \}_{\rm gc} \right) \nonumber \\
   &&+\; F\,\pd{{\bf K}^{\star}}{p_{\|}}\bdot\left[  \left({\bf G}^{\star}\btimes\bhat\right)\; \frac{{\bf B}^{*}}{B_{\|}^{*}}\bdot\nabla f \;-\; \left({\bf F}^{\star}\btimes\bhat\right)\; \frac{{\bf B}^{*}}{B_{\|}^{*}}\bdot\nabla g \right. \nonumber \\
    &&\left.\hspace*{0.3in}+\; {\bf G}^{\star}\btimes\left(  \frac{q}{c}\,\{{\bf X},\, f\}_{\rm gc}\btimes{\bf B}^{*}\right) \;-\; {\bf F}^{\star}\btimes\left(  \frac{q}{c}\,\{{\bf X},\, g\}_{\rm gc}\btimes{\bf B}^{*}\right)\right].
 \end{eqnarray}
 When we combine these expressions into Eq.~\eqref{eq:Maxwell_1_a}, we obtain
 \begin{eqnarray}
 && \int_{X}\nabla\btimes{\bf K}\bdot\frac{\delta^{P}}{\delta{\bf B}}\left[\int_{Z} F_{\rm gc}\,\left({\bf G}^{\star}\bdot\{ {\bf X},\; f\}_{\rm gc} -\frac{}{} {\bf F}^{\star}\bdot\{ {\bf X},\; g\}_{\rm gc} \right) \right] \nonumber \\
  &&=\; \frac{q}{c} \int_{Z} F_{\rm gc} \left[ \left\{X^{i}, K_{i}^{\star}\right\}_{\rm gc}\;\left({\bf G}^{\star}\bdot\{{\bf X},\, f\}_{\rm gc} 
 \;-\frac{}{} {\bf F}^{\star}\bdot\{{\bf X},\, g\}_{\rm gc} \right) \right] \nonumber \\
  &&+\; \frac{q}{c} \int_{Z} F_{\rm gc} \left(\left\{{\bf K}^{\star},\, g\right\}_{\rm gc}\bdot\{{\bf X},{\bf X}\}_{\rm gc}\bdot{\bf F}^{\star} \;+\; {\bf F}^{\star}\bdot\left\{{\bf X},\, {\bf K}^{\star}\right\}_{\rm gc}\bdot\{{\bf X},\, g \}_{\rm gc} \right) \nonumber \\
  &&-\; \frac{q}{c} \int_{Z} F_{\rm gc} \left(\left\{{\bf K}^{\star},\, f\right\}_{\rm gc}\bdot\{{\bf X},{\bf X}\}_{\rm gc}\bdot{\bf G}^{\star} \;+\; {\bf G}^{\star}\bdot\left\{{\bf X},\, {\bf K}^{\star}\right\}_{\rm gc}\bdot\{{\bf X},\, f \}_{\rm gc} \right) \nonumber \\
   &&+\; \int_{Z} F\,\pd{{\bf K}^{\star}}{p_{\|}}\bdot\left[  \left({\bf G}^{\star}\btimes\bhat\right)\; \frac{{\bf B}^{*}}{B_{\|}^{*}}\bdot\nabla f \;-\; \left({\bf F}^{\star}\btimes\bhat\right)\; \frac{{\bf B}^{*}}{B_{\|}^{*}}\bdot\nabla g \;-\;
   \left({\bf G}^{\star}\btimes\nabla f \;-\frac{}{} {\bf F}^{\star}\btimes\nabla g\right) \right. \nonumber \\
    &&\left.\hspace*{0.3in}+\; {\bf G}^{\star}\btimes\left(  \frac{q}{c}\,\{{\bf X},\, f\}_{\rm gc}\btimes{\bf B}^{*}\right) \;-\; {\bf F}^{\star}\btimes\left(  \frac{q}{c}\,\{{\bf X},\, g\}_{\rm gc}\btimes{\bf B}^{*}\right)\right].
  \end{eqnarray}
  We now note that the terms associated with $\partial{\bf K}^{\star}/\partial p_{\|}$ cancel out exactly, while the expression for the Jacobi property involves cyclic permutations of the functionals $({\cal F},{\cal G},{\cal K})$, with the combination
  $[f;{\bf G}^{\star},{\bf K}^{\star}]$:
  \begin{eqnarray}
  && -\,4\pi c \int_{X}\left[\nabla\btimes{\bf K}\bdot\left(\frac{\delta^{P}}{\delta{\bf B}}\int_{Z} F_{\rm gc}\,{\bf G}^{\star}\bdot\{ {\bf X},\; f\}_{\rm gc} \right) - \nabla\btimes{\bf G}\bdot\left(\frac{\delta^{P}}{\delta{\bf B}}\int_{Z} F_{\rm gc}\,{\bf K}^{\star}\bdot
  \{ {\bf X},\; f\}_{\rm gc} \right) \right] \nonumber \\
   &&=\;  4\pi q \int_{Z} F_{\rm gc}\;{\sf Jac}_{M}^{(2)}[f;{\bf G}^{\star},{\bf K}^{\star}],
   \label{eq:M_2}
   \end{eqnarray}
   where the second-order Maxwell sub-bracket contribution is
   \begin{eqnarray}
   {\sf Jac}_{M}^{(2)}[f;{\bf G}^{\star},{\bf K}^{\star}] &=& \left\{ {\bf K}^{\star},\; f\right\}_{\rm gc}\bdot\{{\bf X},{\bf X}\}_{\rm gc}\bdot{\bf G}^{\star} \;-\; \left\{ {\bf G}^{\star},\; f\right\}_{\rm gc}\bdot\{{\bf X},{\bf X}\}_{\rm gc}\bdot{\bf K}^{\star} \nonumber \\
    &&+\;  {\bf G}^{\star}\bdot\left\{{\bf X},\, {\bf K}^{\star}\right\}_{\rm gc}\bdot\{{\bf X},\, f \}_{\rm gc} \;-\; {\bf K}^{\star}\bdot\left\{{\bf X},\, {\bf G}^{\star}\right\}_{\rm gc}\bdot\{{\bf X},\, f \}_{\rm gc} \nonumber \\
    &&+\; \left\{X^{i}, G_{i}^{\star}\right\}_{\rm gc}\;{\bf K}^{\star}\bdot\{{\bf X},\, f\}_{\rm gc}  \;-\; \left\{X^{i}, K_{i}^{\star}\right\}_{\rm gc}\;{\bf G}^{\star}\bdot\{{\bf X},\, f\}_{\rm gc}.
    \label{eq:Maxwell_2_final}
 \end{eqnarray}
 Hence, by combining Eqs.~\eqref{eq:VI_2_final}, \eqref{eq:Jac_2}, and \eqref{eq:Maxwell_2_final}, we finally obtain
\begin{equation}
{\sf Jac}_{VI}^{(2)}[f;{\bf G}^{\star},{\bf K}^{\star}] + {\sf Jac}_{J}^{(2)}[f;{\bf G}^{\star},{\bf K}^{\star}] + {\sf Jac}_{M}^{(2)}[f;{\bf G}^{\star},{\bf K}^{\star}]\;=\; {\sf Jac}_{VM}^{(2)}[f;{\bf G}^{\star},{\bf K}^{\star}],
\label{eq:VM_1}
\end{equation}
where the second-order Vlasov-Maxwell term ${\sf Jac}_{VM}^{(2)}[f;{\bf G}^{\star},{\bf K}^{\star}]$ is defined in Eq.~\eqref{eq:MV_2}.

\subsubsection{Second-order term}

For the second-order term in $4\pi q$ in Eq.~\eqref{eq:gcVM_bracket_app}, we begin with Eq.~\eqref{eq:Maxwell_two}:
  \begin{equation}
-\,4\pi c\int_{X}\left[\nabla\btimes{\bf K}\bdot\frac{\delta^{P}}{\delta{\bf B}}\left(\int_{Z} F_{\rm gc}\,{\bf F}^{\star}\bdot\{ {\bf X},\; {\bf X}\}_{\rm gc}\bdot{\bf G}^{\star}\right) + \leftturn\right] \;=\; 4\pi q\int_{Z} F_{\rm gc}\;{\sf Jac}_{M}^{(3)}[{\bf F}^{\star},
{\bf G}^{\star},{\bf K}^{\star}].
\label{eq:Maxwell_2_a}
\end{equation}
where the third-order Maxwell sub-bracket contribution is
\begin{equation}
{\sf Jac}_{M}^{(3)}[{\bf F}^{\star},{\bf G}^{\star},{\bf K}^{\star}] \;\equiv\; \frac{c}{qB_{\|}^{*}}\left[ \pd{{\bf F}^{\star}}{p_{\|}}\bdot\left({\bf G}^{\star}\btimes{\bf K}^{\star}\right) + \pd{{\bf G}^{\star}}{p_{\|}}\bdot\left({\bf K}^{\star}\btimes{\bf F}^{\star}\right) + 
\pd{{\bf K}^{\star}}{p_{\|}}\bdot\left({\bf F}^{\star} \btimes{\bf G}^{\star}\right)\right].
\label{eq:Jac_M_3}
\end{equation}
We now use the identity
\begin{eqnarray}
\frac{c}{qB_{\|}^{*}}\;\pd{{\bf K}^{\star}}{p_{\|}}\bdot\left({\bf F}^{\star}\btimes{\bf G}^{\star}\right) &=& \left(\pd{{\bf K}^{\star}}{p_{\|}}\btimes\frac{c\bhat}{qB_{\|}^{*}}\right)\bdot\left[ \left({\bf F}^{\star}\btimes{\bf G}^{\star}\right)\btimes\frac{{\bf B}^{*}}{B_{\|}^{*}}\right] \;+\; \frac{c\bhat}{qB_{\|}^{*}}\bdot\left({\bf F}^{\star}\btimes{\bf G}^{\star}\right)\;\frac{{\bf B}^{*}}{B_{\|}^{*}}\bdot\pd{{\bf K}^{\star}}{p_{\|}} \nonumber \\
 &=& \left( {\bf F}^{\star}\bdot\frac{{\bf B}^{*}}{B_{\|}^{*}}\right) \pd{{\bf K}^{\star}}{p_{\|}}\bdot\{{\bf X}, {\bf X}\}_{\rm gc}\bdot{\bf G}^{\star} \;-\; \left( {\bf G}^{\star}\bdot\frac{{\bf B}^{*}}{B_{\|}^{*}}\right)\; \pd{{\bf K}^{\star}}{p_{\|}}\bdot\{{\bf X}, {\bf X}\}_{\rm gc}\bdot{\bf F}^{\star} \nonumber \\
  &&+\; \frac{c\bhat}{qB_{\|}^{*}}\bdot\left({\bf F}^{\star}\btimes{\bf G}^{\star}\right)\;\frac{{\bf B}^{*}}{B_{\|}^{*}}\bdot\pd{{\bf K}^{\star}}{p_{\|}},
\end{eqnarray}
where the last term on the right side is
\[  \frac{c\bhat}{qB_{\|}^{*}}\bdot\left({\bf F}^{\star}\btimes{\bf G}^{\star}\right)\;\frac{{\bf B}^{*}}{B_{\|}^{*}}\bdot\pd{{\bf K}^{\star}}{p_{\|}} \;=\; -\,\left\{ X^{i},\, K_{i}^{\star}\right\}_{\rm gc}\;{\bf F}^{\star}\bdot\{{\bf X}, {\bf X}\}_{\rm gc}\bdot{\bf G}^{\star} \;+\;
\left(\frac{c\bhat}{qB_{\|}^{*}}\bdot\nabla\btimes{\bf K}^{\star}\right) {\bf F}^{\star}\bdot\{{\bf X}, {\bf X}\}_{\rm gc}\bdot{\bf G}^{\star}. \]
Next, we use the identity
\begin{eqnarray}
\left({\bf K}^{\star}\bdot\{{\bf X}, {\bf F}^{\star}\}_{\rm gc} \;-\frac{}{} {\bf F}^{\star}\bdot\{{\bf X}, {\bf K}^{\star}\}_{\rm gc}\right)\bdot\{{\bf X},{\bf X}\}_{\rm gc}\bdot{\bf G}^{\star} &=& \left[\left({\bf F}^{\star}\bdot\frac{{\bf B}^{*}}{B_{\|}^{*}}\right)
\pd{{\bf K}^{\star}}{p_{\|}} - \left({\bf K}^{\star}\bdot\frac{{\bf B}^{*}}{B_{\|}^{*}}\right)\pd{{\bf F}^{\star}}{p_{\|}}\right]\bdot{\bf G}^{\star}\btimes\frac{c\bhat}{qB_{\|}^{*}} \\
 &&+\; \left({\bf F}^{\star}\btimes\frac{c\bhat}{qB_{\|}^{*}} \bdot\nabla{\bf K}^{\star}  - {\bf K}^{\star}\btimes\frac{c\bhat}{qB_{\|}^{*}} \bdot\nabla{\bf F}^{\star}\right)\bdot{\bf G}^{\star}\btimes\frac{c\bhat}{qB_{\|}^{*}}, \nonumber 
\end{eqnarray}
so that Eq.~\eqref{eq:Jac_M_3} yields the final expression for the third-order Maxwell sub-bracket contribution
\begin{equation}
{\sf Jac}_{M}^{(3)}[{\bf F}^{\star},{\bf G}^{\star},{\bf K}^{\star}] \;=\; -\;{\sf Jac}_{J}^{(3)}[{\bf F}^{\star},{\bf G}^{\star},{\bf K}^{\star}] \;-\; \left[ \left({\bf K}^{\star}\bdot\{{\bf X}, {\bf F}^{\star}\}_{\rm gc} \;-\frac{}{} {\bf F}^{\star}\bdot\{{\bf X}, {\bf K}^{\star}\}_{\rm gc}\right)\bdot\{{\bf X},{\bf X}\}_{\rm gc}\bdot{\bf G}^{\star} + \leftturn \right],
\label{eq:Maxwell_3_final}
 \end{equation}
 where ${\sf Jac}_{J}^{(3)}[{\bf F}^{\star},{\bf G}^{\star},{\bf K}^{\star}]$ is given by Eq.~\eqref{eq:Jac_3} and we used the identity
\[ {\bf F}^{\star}\btimes\frac{c\bhat}{qB_{\|}^{*}} \bdot\nabla{\bf K}^{\star}\bdot{\bf G}^{\star}\btimes\frac{c\bhat}{qB_{\|}^{*}} - {\bf G}^{\star}\btimes\frac{c\bhat}{qB_{\|}^{*}} \bdot\nabla{\bf K}^{\star}\bdot{\bf F}^{\star}\btimes\frac{c\bhat}{qB_{\|}^{*}} \;=\;
\left(\frac{c\bhat}{qB_{\|}^{*}}\bdot\nabla\btimes{\bf K}^{\star}\right) {\bf F}^{\star}\bdot\{{\bf X},{\bf X}\}_{\rm gc}\bdot{\bf G}^{\star}. \]
 Hence, by combining Eqs.~\eqref{eq:VI_3_final}, \eqref{eq:Jac_3}, and \eqref{eq:Maxwell_3_final}, we finally obtain
\begin{equation}
{\sf Jac}_{VI}^{(3)}[{\bf F}^{\star},{\bf G}^{\star},{\bf K}^{\star}]  + {\sf Jac}_{J}^{(3)}[{\bf F}^{\star},{\bf G}^{\star},{\bf K}^{\star}] + {\sf Jac}_{M}^{(3)}[{\bf F}^{\star},{\bf G}^{\star},{\bf K}^{\star}] \;=\; {\sf Jac}_{VM}^{(3)}[{\bf F}^{\star},{\bf G}^{\star},{\bf K}^{\star}],
\label{eq:VM_3}
\end{equation}
where the third-order Vlasov-Maxwell term  ${\sf Jac}_{VM}^{(3)}[f;{\bf G}^{\star},{\bf K}^{\star}]$ is defined in Eq.~\eqref{eq:MV_3}.

\section{Proof of the Jacobi property}

By combining the Vlasov-Maxwell contributions \eqref{eq:Vlasov_0}, \eqref{eq:MV_1}, \eqref{eq:MV_2}, and \eqref{eq:MV_3}, the Jacobi property \eqref{eq:Jacobi} of the guiding-center Vlasov-Maxwell bracket \eqref{eq:gcVM_bracket} can now be expressed as a series in powers of $(4\pi q)$ up to third order:
\begin{eqnarray}
{\cal Jac}[{\cal F},{\cal G},{\cal K}]  &=& \int_{Z} F_{\rm gc}\;{\sf Jac}_{VM}^{(0)}[f,g,k] \;+\; 4\pi q \int_{Z} F_{\rm gc} \left( {\sf Jac}_{VM}^{(1)}[f,g; {\bf K}^{\star}] + \leftturn \right) \nonumber \\
  &&+\; (4\pi q)^{2} \int_{Z} F_{\rm gc} \left( {\sf Jac}_{VM}^{(2)}[f;{\bf G}^{\star},{\bf K}^{\star}] + \leftturn \right) +  (4\pi q)^{3} \int_{Z} F_{\rm gc}\;{\sf Jac}_{VM}^{(3)}[{\bf F}^{\star},{\bf G}^{\star},{\bf K}^{\star}] \nonumber \\
 &=&  \int_{Z} F_{\rm gc}\left( \left\{ \{ f,\; g\}_{\rm gc},\frac{}{} k \right\}_{\rm gc} + \left\{ \{ g,\; k\}_{\rm gc},\frac{}{} f \right\}_{\rm gc}  + \left\{ \{ k,\; f\}_{\rm gc},\frac{}{} g \right\}_{\rm gc} \right) 
\label{eq:Jacobi_4piq} \\
 &&+ 4\pi q \int_{Z} F_{\rm gc} \left[ F_{i}^{\star} \left( \left\{ \{ X^{i},\; g\}_{\rm gc},\frac{}{} k \right\}_{\rm gc} + \left\{ \{ g,\; k\}_{\rm gc},\frac{}{} X^{i} \right\}_{\rm gc}  + \left\{ \{ k,\; X^{i}\}_{\rm gc},\frac{}{} g \right\}_{\rm gc} \right) \;+\; \leftturn \right]  \nonumber \\
 &&+ (4\pi q)^{2} \int_{Z} F_{\rm gc} \left[ F_{i}^{\star}\,G_{j}^{\star} \left( \left\{ \{ X^{i},\; X^{j}\}_{\rm gc},\frac{}{} k \right\}_{\rm gc} + \left\{ \{ X^{j},\; k\}_{\rm gc},\frac{}{} X^{i} \right\}_{\rm gc}  + \left\{ \{ k,\; X^{i}\}_{\rm gc},
 \frac{}{} X^{j} \right\}_{\rm gc} \right) \;+\; \leftturn \right] \nonumber \\
  &&+ (4\pi q)^{3} \int_{Z} F_{\rm gc} \left[ F_{i}^{\star}\,G_{j}^{\star}\,K_{\ell}^{\star} \left( \left\{ \{ X^{i}, X^{j}\}_{\rm gc},\frac{}{} X^{\ell} \right\}_{\rm gc} + \left\{ \{ X^{j}, X^{\ell}\}_{\rm gc},\frac{}{} X^{i} 
  \right\}_{\rm gc} + \left\{ \{ X^{\ell}, X^{i}\}_{\rm gc},\frac{}{} X^{j} \right\}_{\rm gc} \right) \right], \nonumber
\end{eqnarray}
where all additional terms have cancelled out exactly as shown in Sec.~\ref{sec:Jacobi}. In Eq.~\eqref{eq:Jacobi_4piq}, it is clear that the Jacobi property of the guiding-center Vlasov-Maxwell bracket \eqref{eq:gcVM_bracket} is inherited from the Jacobi property \eqref{eq:gcPB_Jacobi} of the guiding-center Poisson bracket \eqref{eq:PB_gc}, since each term in Eq.~\eqref{eq:Jacobi_4piq} vanishes identically because of this latter property. Hence, the Jacobi property for the guiding-center Vlasov-Maxwell bracket \eqref{eq:gcVM_bracket} holds under the condition \eqref{eq:div_Bstar}, which holds according to Eq.~\eqref{eq:B*_def}.

\section{Summary}

The proof of the Jacobi property of a functional bracket that forms the basis for the Hamiltonian structure of reduced Vlasov-Maxwell equations is often a challenging task (see, for example, Ref.~\cite{Brizard_2016}). While we can also appeal to rigorous theoretical grounds for the validity of the Jacobi property of the guiding-center Vlasov-Maxwell bracket \eqref{eq:gcVM_bracket}, an explicit proof of Eq.~\eqref{eq:Jacobi} was presented in these notes. As expected, the Jacobi property is inherited from the Jacobi property \eqref{eq:gcPB_Jacobi} of the guiding-center Poisson bracket \eqref{eq:PB_gc}.

\appendix

\section{Poisson-bracket Identities}

In this Appendix, we derive several identities associated with the guiding-center Poisson bracket \eqref{eq:PB_gc}. First, we have the integral identity
\begin{equation}
\int_{Z} B_{\|}^{*}\,\{ f,\; g\}_{\rm gc} \;=\; \int_{Z} \pd{}{Z^{\alpha}}\left( B_{\|}^{*}\,f\;\left\{ Z^{\alpha},\; g\right\}_{\rm gc}\right) \;=\; 0,
\label{eq:Id_gc_0}
\end{equation}
which holds for any functions $(f,g)$. This identity yields the formula $\int_{Z} B_{\|}^{*}\{f, g\}_{\rm gc}\,k = \int_{Z} B_{\|}^{*}\,f\,\{g, k\}_{\rm gc}$ after integration by parts is performed.

Next, for any arbitrary vector-valued function ${\bf R}$, we find the divergence identity
\begin{equation}
\left\{ X^{i},\frac{}{} R_{i}\right\}_{\rm gc} \;=\; \frac{c\bhat}{qB_{\|}^{*}}\bdot\nabla\btimes{\bf R} \;+\; \frac{{\bf B}^{*}}{B_{\|}^{*}}\bdot\pd{\bf R}{p_{\|}},
\label{eq:Id_gc_Jac}
\end{equation}
where summation over repeated indices on the left side is implied. In addition, for an arbitrary function $f$, we find 
\begin{equation}
\{{\bf R},\,f\}_{\rm gc} \;=\; \{{\bf X},\, f\}_{\rm gc}\bdot\nabla{\bf R} \;-\; \left(\frac{{\bf B}^{*}}{B_{\|}^{*}}\bdot\nabla f\right)\,\pd{\bf R}{p_{\|}},
\label{eq:Id_gc_Rf}
\end{equation}
where $\{{\bf X}, f\}_{\rm gc} = ({\bf B}^{*}/B_{\|}^{*})\,\partial f/\partial p_{\|} + (c\bhat/qB_{\|}^{*})\btimes\nabla f$.

Lastly, for two arbitrary functions $(f,g)$, we find the Poisson-bracket identity
\begin{equation}
\left(\frac{c}{q}\right)\,\nabla f\btimes\nabla g \;=\; \left\{{\bf X},\,g\right\}_{\rm gc}\;{\bf B}^{*}\bdot\nabla f \;-\;  \left\{{\bf X},\,f\right\}_{\rm gc}\;{\bf B}^{*}\bdot\nabla g \;-\; {\bf B}^{*}\;\{f,\, g\}_{\rm gc},
\label{eq:Ig_gc_fxg}
\end{equation} 
which is derived from the identity
\[ \bhat\btimes\left[{\bf B}^{*}\frac{}{}\btimes\left(\nabla f\btimes\frac{}{}\nabla g\right)\right] \;=\; \left\{ \begin{array}{l}
\bhat\bdot(\nabla f\btimes\nabla g)\,{\bf B}^{*} \;-\; B_{\|}^{*}\;\nabla f\btimes\nabla g \\
\\
\bhat\btimes\nabla f\;({\bf B}^{*}\bdot\nabla g) \;-\; \bhat\btimes\nabla g\;({\bf B}^{*}\bdot\nabla f)
\end{array} \right. \]
Next, we use the identity
\begin{eqnarray}
\{ {\bf X}, f\}_{\rm gc} \btimes\{ {\bf X}, g\}_{\rm gc} &=& \left(\frac{{\bf B}^{*}}{B_{\|}^{*}}\,\pd{f}{p_{\|}} + \frac{c\bhat}{qB_{\|}^{*}}\btimes\nabla f\right)\btimes \left(\frac{{\bf B}^{*}}{B_{\|}^{*}}\,\pd{g}{p_{\|}} + \frac{c\bhat}{qB_{\|}^{*}}\btimes\nabla g\right)
\nonumber \\
 &=& \frac{{\bf B}^{*}}{B_{\|}^{*}}\btimes\left[ \pd{f}{p_{\|}} \left(\frac{c\bhat}{qB_{\|}^{*}}\btimes\nabla g\right) - \pd{g}{p_{\|}} \left(\frac{c\bhat}{qB_{\|}^{*}}\btimes\nabla f\right)\right] \;+\; \left(\frac{c\bhat}{qB_{\|}^{*}}\btimes\nabla f\right)\btimes
 \left(\frac{c\bhat}{qB_{\|}^{*}}\btimes\nabla g\right) \nonumber \\
  &=& \frac{c}{qB_{\|}^{*}} \left[  \left(\nabla f\,\pd{g}{p_{\|}} - \pd{f}{p_{\|}}\,\nabla g\right) \;-\; \bhat\;\{ f,\; g\}_{\rm gc} \right],
\label{eq:Id_gc_curl}
\end{eqnarray}
to obtain
\begin{eqnarray}
\nabla\btimes{\bf R}\bdot\left(\{ {\bf X}, f\}_{\rm gc}\frac{}{}\btimes\{ {\bf X}, g\}_{\rm gc}\right) &=& \frac{c}{qB_{\|}^{*}} \left[ \nabla\btimes{\bf R}\bdot\left(\nabla f\,\pd{g}{p_{\|}} - \pd{f}{p_{\|}}\,\nabla g\right) \;-\; \bhat\bdot\nabla\btimes{\bf R}\;\{ f,\; g\}_{\rm gc} \right] \nonumber \\
 &=& \{{\bf R}, f\}_{\rm gc}\bdot\{{\bf X},g\}_{\rm gc} \;-\; \{{\bf R}, g\}_{\rm gc}\bdot\{{\bf X},f\}_{\rm gc} \nonumber \\
  &&+\; \pd{\bf R}{p_{\|}}\bdot\left[ \{{\bf X},\,g\}_{\rm gc} \left(\frac{{\bf B}^{*}}{B_{\|}^{*}}\bdot\nabla f\right) \;-\; \{{\bf X},\,f\}_{\rm gc} \left(\frac{{\bf B}^{*}}{B_{\|}^{*}}\bdot\nabla g\right) \right]. 
  \label{eq:Id_gc_Maxwell_1}
\end{eqnarray}

\section{Magnetic Functional Derivatives in the Maxwell Sub-Bracket}

In this Appendix, we calculate the magnetic functional derivatives in the Maxwell sub-bracket that appear in Eq.~\eqref{eq:Jac_fg-k}. These functional derivatives are separated into Jacobian and $n^{th}$-order $(n = 0,1,2)$ derivatives.

\subsection{Jacobian contributions}

Each term in the Poisson functional derivative \eqref{eq:gcVM_double} includes a guiding-center Poisson bracket, which contains the Jacobian $B_{\|}^{*}$ as a denominator. Hence, the double-bracket expression \eqref{eq:Jac_fg-k} includes the Jacobian contribution
\begin{equation}
\delta_{J}^{P}\left[{\cal F},\frac{}{}{\cal G}\right]_{\rm gc} \;\equiv\; -\; \int_{Z} \frac{F_{\rm gc}}{B_{\|}^{*}}\;\left[{\cal F},\frac{}{}{\cal G}\right]_{F_{\rm gc}}\;\delta B_{\|}^{*},
\label{eq:delta_Jac}
\end{equation}
where $\left[{\cal F},\frac{}{}{\cal G}\right]_{F_{\rm gc}} \equiv \delta^{P}[{\cal F},{\cal G}]_{\rm gc}/\delta F_{\rm gc}$ is given by Eq.~\eqref{eq:gcVM_double}, and
\begin{eqnarray}
\delta B_{\|}^{*} &=& \delta\bhat\bdot{\bf B}^{*} \;+\; \bhat\bdot\delta{\bf B}^{*} \;=\; \left(\delta{\bf B}\bdot\frac{q}{c}\,\pd{\mathbb{P}_{\|}}{p_{\|}}\right)\bdot{\bf B}^{*} \;+\; \bhat\bdot\left[ \delta{\bf B} \;+\; \nabla\btimes\left(\mathbb{P}_{\|}\bdot\delta{\bf B}\right) \right]
\nonumber \\
 &=& \bhat\bdot\delta{\bf B} \;+\; \nabla\bdot\left[\left(\mathbb{P}_{\|}\bdot\delta{\bf B}\right) \btimes\bhat\right] \;+\; \pd{}{p_{\|}}\left[\delta{\bf B}\bdot\left(\frac{q}{c}\,\mathbb{P}_{\|}\bdot{\bf B}^{*}\right) \right],
 \label{eq:deltaB_star}
\end{eqnarray}
so that, for an arbitrary guiding-center function $w$, we find (after integrating by parts)
\[ \int_{Z} w\,\delta B_{\|}^{*} \;=\; \int_{X} \delta{\bf B}\bdot\int_{P} \left[ w\,\bhat \;-\; \frac{q}{c}\,\mathbb{P}_{\|}\bdot\left({\bf B}^{*}\,\pd{w}{p_{\|}} \;+\; \frac{c\bhat}{q}\btimes\nabla w \right) \right] \;\equiv\;  \int_{X} \delta{\bf B}\bdot\int_{P} \left( w\,\bhat \;-\; \frac{qB_{\|}^{*}}{c}\,\mathbb{P}_{\|}\bdot\left\{ {\bf X},\frac{}{} w \right\}_{\rm gc} \right). \]
For an arbitrary vector field ${\bf K}$, we obtain
\begin{eqnarray}
4\pi c\int_{X} \nabla\btimes{\bf K}\bdot\int_{P} w\,\fd{B_{\|}^{*}}{\bf B} &=& 4\pi q\int_{Z} \left[w\,\frac{c\bhat}{q}\bdot\nabla\btimes{\bf K} \;-\; B_{\|}^{*}\,\left(\mathbb{P}_{\|}\bdot\nabla\btimes{\bf K}\right) \bdot\left\{ {\bf X},\frac{}{} w \right\}_{\rm gc} \right] 
\nonumber \\
 &=& 4\pi q \int_{Z} B_{\|}^{*} \left( \left\{ X^{i},\frac{}{} w\,K_{i}\right\}_{\rm gc} - {\bf K}^{\star}\bdot\left\{ {\bf X},\frac{}{} w \right\}_{\rm gc} \right) \equiv -\,4\pi q \int_{Z} B_{\|}^{*}\,{\bf K}^{\star}\bdot\left\{ {\bf X},\frac{}{} w \right\}_{\rm gc},
 \end{eqnarray}
 where we used the identities \eqref{eq:Id_gc_0}-\eqref{eq:Id_gc_Jac} and the definition ${\bf K}^{\star} = {\bf K} + \mathbb{P}_{\|}\bdot\nabla\btimes{\bf K}$. Integrating by parts, using the Poisson-bracket Leibnitz formula ${\bf K}^{\star}\bdot\{{\bf X}, w\}_{\rm gc} = \{X^{i},\,
 w K_{i}^{\star}\}_{\rm gc} - \{ X^{i}, K_{i}^{\star}\}_{\rm gc}\,w$, we finally obtain the series expansion in powers of $4\pi q$:
 \begin{eqnarray}
 4\pi c\int_{X} \nabla\btimes{\bf K}\bdot\int_{P} w\,\fd{B_{\|}^{*}}{\bf B} &=& \int_{Z} F_{\rm gc} \left\{ X^{i}, K_{i}^{\star}\right\}_{\rm gc} \left[ 4\pi q\;\{ f, g\}_{\rm gc} \;-\frac{}{} (4\pi q)^{2} \left( {\bf F}^{\star}\bdot\{{\bf X}, g\}_{\rm gc} \;-\frac{}{} {\bf G}^{\star}\bdot
 \{{\bf X}, f\}_{\rm gc} \right) \right. \nonumber \\
  &&\left.+\; (4\pi q)^{3}\frac{}{} \left({\bf F}^{\star}\bdot\{{\bf X}, \frac{}{} {\bf X}\}_{\rm gc}\bdot{\bf G}^{\star}\right) \right],
  \label{eq:Maxwell_Jac}
 \end{eqnarray}
after substituting $w \equiv [{\cal F},{\cal G}]_{F_{\rm gc}}\,F_{\rm gc}/B_{\|}^{*}$.

\subsection{Zeroth-order Maxwell sub-bracket}

In the Maxwell sub-bracket appearing in Eq.~\eqref{eq:gcVM_bracket_app}, we begin with the magnetic variation of the zeroth-order term
\begin{eqnarray*}
\int_{Z} F \;\delta^{P}\left(B_{\|}^{*}\,\{ f,\; g\}_{\rm gc}\right) &=& \int_{Z} F \left[ \delta{\bf B}^{*}\bdot\left(\nabla f\,\pd{g}{p_{\|}} - \pd{f}{p_{\|}}\right) - \frac{c}{q}\delta\bhat\bdot\nabla f\btimes\nabla g \right] \\
 &=& \int_{Z}\delta{\bf B} \bdot\left[ F \left(\nabla f\,\pd{g}{p_{\|}} - \pd{f}{p_{\|}}\right) + \mathbb{P}_{\|}\bdot\nabla F\btimes\left(\nabla f\,\pd{g}{p_{\|}} - \pd{f}{p_{\|}}\right) + \pd{F}{p_{\|}}\;\mathbb{P}_{\|}\bdot (\nabla f\btimes\nabla g) \right],
 \end{eqnarray*}
where $\delta{\bf B}^{*}$ and $\delta\bhat$ are given in Eq.~\eqref{eq:deltaB_bhat}, and integration by parts were carried out in order to release $\delta{\bf B}$. Hence, we obtain the magnetic functional derivative
\begin{equation} 
\frac{\delta^{P}}{\delta{\bf B}}\left(\int_{Z} F_{\rm gc}\,\{ f,\; g\}_{\rm gc}\right) \;=\; \int_{P}F \left(\nabla f\,\pd{g}{p_{\|}} - \pd{f}{p_{\|}}\right)  \;+\; \mathbb{P}_{\|}\bdot\left[ \pd{F}{p_{\|}}\,(\nabla f\btimes\nabla g) - \left(\nabla f\,\pd{g}{p_{\|}} - \pd{f}{p_{\|}}\,
\nabla g\right)\btimes\nabla F \right].
\label{eq:deltaB_Vlasov_1}
\end{equation}
We now insert this functional derivative in Eq.~\eqref{eq:Jac_fg-k} to obtain
 \begin{eqnarray}
\int_{X}\nabla\btimes{\bf K}\bdot\frac{\delta^{P}}{\delta{\bf B}}\left(\int_{Z} F_{\rm gc}\,\{ f,\; g\}_{\rm gc}\right)  &=& \int_{Z} F\;\nabla\btimes{\bf K}\bdot\left(\nabla f\,\pd{g}{p_{\|}} - \pd{f}{p_{\|}}\right) \nonumber \\
  &&+\; \int_{Z} \left({\bf K}^{\star} - {\bf K}\right)\bdot\left[ \pd{F}{p_{\|}}\,(\nabla f\btimes\nabla g) - \left(\nabla f\,\pd{g}{p_{\|}} - \pd{f}{p_{\|}}\,\nabla g\right)\btimes\nabla F \right],
  \end{eqnarray}
 where we used $\mathbb{P}_{\|}\bdot\nabla\btimes{\bf K} = {\bf K}^{\star} - {\bf K}$. Next, we integrate by parts the terms
 \begin{eqnarray*}
 \int_{Z} \left({\bf K}^{\star} - {\bf K}\right)\bdot\pd{F}{p_{\|}}\,(\nabla f\btimes\nabla g) &=& -\; \int_{Z} F \left[ \pd{{\bf K}^{\star}}{p_{\|}}\bdot(\nabla f\btimes\nabla g) \;+\; \left({\bf K}^{\star} - {\bf K}\right)\bdot\pd{}{p_{\|}}(\nabla f\btimes\nabla g)\right], \\
 \int_{Z} \nabla F\bdot\left(\nabla f\,\pd{g}{p_{\|}} - \pd{f}{p_{\|}}\,\nabla g\right)\btimes\left({\bf K}^{\star} - {\bf K}\right) &=& \int_{Z} F\;\nabla\btimes\left({\bf K}^{\star} - {\bf K}\right)\bdot\left(\nabla f\,\pd{g}{p_{\|}} - \pd{f}{p_{\|}}\,\nabla g\right) \\
  &&+\; \int_{Z} F\; \left({\bf K}^{\star} - {\bf K}\right)\bdot\pd{}{p_{\|}}(\nabla f\btimes\nabla g),
 \end{eqnarray*}
so that, after cancellations, we obtain the final expression to be used in Eq.~\eqref{eq:Maxwell_1}:
\begin{equation}
\int_{X}\nabla\btimes{\bf K}\bdot\frac{\delta^{P}}{\delta{\bf B}}\left(\int_{Z} F_{\rm gc}\,\{ f,\; g\}_{\rm gc}\right) = \int_{Z} F \left[ \nabla\btimes{\bf K}^{\star}\bdot\left( \nabla f\,\pd{g}{p_{\|}} - \pd{f}{p_{\|}}\,\nabla g\right) \;-\; \pd{{\bf K}^{\star}}{p_{\|}}\bdot
(\nabla f\btimes\nabla g) \right].
\label{eq:Maxwell_zero}
\end{equation}

\subsection{First-order Maxwell sub-bracket}

In the Maxwell sub-bracket appearing in Eq.~\eqref{eq:gcVM_bracket_app}, we begin with the magnetic variation of the first-order term
\begin{eqnarray*}
\int_{Z} F \delta^{P}\left(B_{\|}^{*}\,{\bf G}^{\star}\bdot\{ {\bf X},\; f\}_{\rm gc}\right) &=& \int_{Z} F \left[ \delta^{P}{\bf G}^{\star}\bdot B_{\|}^{*}\{ {\bf X},\; f\}_{\rm gc} \;+\; {\bf G}^{\star}\bdot\delta{\bf B}^{*}\,\pd{f}{p_{\|}} + \frac{c}{q}\delta\bhat\bdot\nabla f\btimes
{\bf G}^{\star} \right] \\
 &=& \int_{X} \delta{\bf B}\bdot \int_{P}\left[ F_{\rm gc}\;\frac{\delta^{P}{\bf G}^{\star}}{\delta{\bf B}}\bdot \{ {\bf X},\; f\}_{\rm gc} \;-\; F\;\mathbb{P}_{\|}\bdot\left(\nabla f\btimes\pd{{\bf G}^{\star}}{p_{\|}} \right) \right] \\
  &&+\; \int_{X} \delta{\bf B}\bdot \int_{P}\left[ F\;\pd{f}{p_{\|}}\left({\bf G}^{\star} + \mathbb{P}_{\|}\bdot\nabla\btimes{\bf G}^{\star}\right) + \mathbb{P}_{\|}\bdot\left(\nabla F\,\pd{f}{p_{\|}} - \pd{F}{p_{\|}}\,\nabla f\right)\btimes{\bf G}^{\star} \right],
 \end{eqnarray*}
where we integrated by parts in order to release $\delta{\bf B}$ and the Poisson variation $\delta^{P}{\bf G}^{\star}$ is defined as
\begin{equation}
 \delta^{P}{\bf G}^{\star} \;=\; \delta\mathbb{P}_{\|}\bdot\nabla\btimes{\bf G} \;=\; -\,\frac{\delta{\bf B}}{B}\bdot\left[  \mathbb{P}_{\|}\,(\bhat\bdot\nabla\btimes{\bf G}) + \bhat\,(\mathbb{P}_{\|}\bdot\nabla\btimes{\bf G}) + (\mathbb{P}_{\|}\bdot\nabla\btimes
 {\bf G})\,\bhat \right] \;\equiv\; \delta{\bf B}\bdot\frac{\delta^{P}{\bf G}^{\star}}{\delta{\bf B}}.
 \label{eq:delta_G_B}
 \end{equation}
Hence, we obtain the magnetic functional derivative
 \begin{eqnarray} 
\frac{\delta^{P}}{\delta{\bf B}}\left(\int_{Z} F_{\rm gc}\,{\bf G}^{\star}\bdot\{ {\bf X},\; f\}_{\rm gc}\right) &=& \int_{P}\left[ F_{\rm gc}\;\frac{\delta^{P}{\bf G}^{\star}}{\delta{\bf B}}\bdot \{ {\bf X},\; f\}_{\rm gc} \;-\; F\;\mathbb{P}_{\|}\bdot\left(\nabla f\btimes
\pd{{\bf G}^{\star}}{p_{\|}} \right) \right] \nonumber \\
 &&+\; \int_{P}\left[ F\;\pd{f}{p_{\|}}\left({\bf G}^{\star} + \mathbb{P}_{\|}\bdot\nabla\btimes{\bf G}^{\star}\right) + \mathbb{P}_{\|}\bdot\left(\nabla F\,\pd{f}{p_{\|}} - \pd{F}{p_{\|}}\,\nabla f\right)\btimes{\bf G}^{\star} \right] .
\label{eq:deltaB_Vlasov_2}
\end{eqnarray}
We now insert this functional derivative in Eq.~\eqref{eq:Jac_fg-k} to obtain
 \begin{eqnarray}
\int_{X}\nabla\btimes{\bf K}\bdot\frac{\delta^{P}}{\delta{\bf B}}\left(\int_{Z} F_{\rm gc}\,{\bf G}^{\star}\bdot\{ {\bf X},\; f\}_{\rm gc}\right) &=& \int_{Z} \left[ F_{\rm gc}\;\nabla\btimes{\bf K}\bdot\frac{\delta^{P}{\bf G}^{\star}}{\delta{\bf B}}\bdot\{ {\bf X},\; f\}_{\rm gc}
+ F\,\pd{f}{p_{\|}}\,\nabla\btimes{\bf K}\bdot\left({\bf G}^{\star} + \mathbb{P}_{\|}\bdot\nabla\btimes{\bf G}^{\star}\right)\right] \nonumber \\
 &&+ \int_{Z}  \left({\bf K}^{\star} - {\bf K}\right)\bdot\left[ \left(\pd{f}{p_{\|}}\,\nabla F - \pd{F}{p_{\|}}\,\nabla f\right)\btimes{\bf G}^{\star} - F\,\left(\nabla f\btimes\pd{{\bf G}^{\star}}{p_{\|}}\right) \right].
\end{eqnarray}
Next, we integrate by parts the terms
 \begin{eqnarray*}
\int_{Z}  \left({\bf K}^{\star} - {\bf K}\right)\bdot\left(\pd{f}{p_{\|}}\,\nabla F - \pd{F}{p_{\|}}\,\nabla f\right)\btimes{\bf G}^{\star} &=& \int_{Z} F\,\pd{f}{p_{\|}}\left[{\bf G}^{\star}\bdot\nabla\btimes{\bf K}^{\star} \;-\frac{}{} \nabla\btimes{\bf K}\bdot\left({\bf G}^{\star} + \mathbb{P}_{\|}\bdot\nabla\btimes{\bf G}^{\star}\right) \right] \\
 &&+\; \int_{Z} F\,\nabla f\bdot\left[ \pd{{\bf G}^{\star}}{p_{\|}}\btimes \left({\bf K}^{\star} - {\bf K}\right) \;+\; {\bf G}^{\star}\btimes\pd{{\bf K}^{\star}}{p_{\|}}\right]
\end{eqnarray*}
so that, after cancellations, we obtain 
 \begin{eqnarray*}
\int_{X}\nabla\btimes{\bf K}\bdot\frac{\delta^{P}}{\delta{\bf B}}\left(\int_{Z} F_{\rm gc}\,{\bf G}^{\star}\bdot\{ {\bf X},\; f\}_{\rm gc}\right) &=& \int_{Z} F_{\rm gc}\;\nabla\btimes{\bf K}\bdot\frac{\delta^{P}{\bf G}^{\star}}{\delta{\bf B}}\bdot\{ {\bf X},\; f\}_{\rm gc} \\
 &&+\; \int_{Z} F\,{\bf G}^{\star}\bdot\left( \pd{f}{p_{\|}}\,\nabla\btimes{\bf K}^{\star} - \nabla f\btimes\pd{{\bf K}^{\star}}{p_{\|}}\right),
 \end{eqnarray*}
 which yields the final expression to be used in Eq.~\eqref{eq:Maxwell_1_a}:
 \begin{eqnarray}
 \int_{X}\nabla\btimes{\bf K}\bdot\frac{\delta^{P}}{\delta{\bf B}}\left[\int_{Z} F_{\rm gc}\,\left({\bf G}^{\star}\bdot\{ {\bf X},\; f\}_{\rm gc} -\frac{}{} {\bf F}^{\star}\bdot\{ {\bf X},\; g\}_{\rm gc} \right) \right] &=& \int_{Z} F\,\nabla\btimes{\bf K}^{\star}\bdot\left( {\bf G}^{\star}\;\pd{f}{p_{\|}} \;-\; {\bf F}^{\star}\;\pd{g}{p_{\|}} \right) \nonumber \\
  &&-\; \int_{Z} F\;\pd{{\bf K}^{\star}}{p_{\|}}\bdot\left({\bf G}^{\star}\btimes\nabla f \;-\frac{}{} {\bf F}^{\star}\btimes\nabla g\right),
  \label{eq:Maxwell_one}
\end{eqnarray}
where we have used Eq.~\eqref{eq:delta_G_B} to obtain the identity
 \begin{eqnarray}
B\left( \nabla\btimes{\bf K}\bdot\frac{\delta^{P}{\bf G}^{\star}}{\delta{\bf B}} \;-\; \nabla\btimes{\bf G}\bdot\frac{\delta^{P}{\bf K}^{\star}}{\delta{\bf B}}\right) &=& \nabla\btimes{\bf G}\bdot\left[ \mathbb{P}_{\|}\,(\bhat\bdot\nabla\btimes{\bf K}) \;+\; \bhat\;
\left(\mathbb{P}_{\|}\bdot\nabla\btimes{\bf K}\right) \;+\; \left(\mathbb{P}_{\|}\bdot\nabla\btimes{\bf K}\right)\;\bhat \right] 
\label{eq:Id_delta_B} \\
 &&-\; \nabla\btimes{\bf K}\bdot\left[ \mathbb{P}_{\|}\,(\bhat\bdot\nabla\btimes{\bf G}) \;+\; \bhat\;
\left(\mathbb{P}_{\|}\bdot\nabla\btimes{\bf G}\right) \;+\; \left(\mathbb{P}_{\|}\bdot\nabla\btimes{\bf G}\right)\;\bhat \right] = 0.
\nonumber
\end{eqnarray}
Hence, the magnetic functional derivatives $(\delta^{P}{\bf F}^{\star}/\delta{\bf B},\cdots)$ do not contribute to the Maxwell sub-bracket and the Jacobi property \eqref{eq:Jacobi}, which may be viewed as an extension of the Bracket theorem.

\subsection{Second-order Maxwell sub-bracket}

In the Maxwell sub-bracket appearing in Eq.~\eqref{eq:gcVM_bracket_app}, we begin with the magnetic variation of the second-order term
\begin{eqnarray*}
\int_{Z} F \delta^{P}\left(B_{\|}^{*}\,{\bf F}^{\star}\bdot\{ {\bf X},\; {\bf X}\}_{\rm gc}\bdot{\bf G}^{\star}\right) &=& -\,\int_{Z} F \left[ \frac{c}{q}\,\delta\bhat\bdot{\bf F}^{\star}\btimes{\bf G}^{\star} \;+\; \delta^{P}{\bf F}^{\star}\bdot \left({\bf G}^{\star}\btimes
\frac{c\bhat}{q} \right) \;-\;  \delta^{P}{\bf G}^{\star}\bdot \left({\bf F}^{\star}\btimes\frac{c\bhat}{q} \right) \right] \\
 &=& -\;\int_{X}\delta{\bf B}\bdot\int_{P} F \left[ \pd{\mathbb{P}_{\|}}{p_{\|}}\bdot{\bf F}^{\star}\btimes{\bf G}^{\star} \;+\; \frac{\delta^{P}{\bf F}^{\star}}{\delta{\bf B}}\bdot \left({\bf G}^{\star}\btimes
\frac{c\bhat}{q} \right) \;-\;  \frac{\delta^{P}{\bf G}^{\star}}{\delta{\bf B}}\bdot \left({\bf F}^{\star}\btimes\frac{c\bhat}{q} \right) \right]
 \end{eqnarray*}
 which yields the final expression to be used in Eq.~\eqref{eq:Maxwell_2_a}:
  \begin{eqnarray}
 &&\int_{X}\left[\nabla\btimes{\bf K}\bdot\frac{\delta^{P}}{\delta{\bf B}}\left(\int_{Z} F_{\rm gc}\,{\bf F}^{\star}\bdot\{ {\bf X},\; {\bf X}\}_{\rm gc}\bdot{\bf G}^{\star}\right) + \leftturn\right] \nonumber \\
  &&\hspace*{0.5in}=\; -\,\int_{Z} F \left[ \pd{{\bf F}^{\star}}{p_{\|}}\bdot\left({\bf G}^{\star}\btimes{\bf K}^{\star}\right) + \pd{{\bf G}^{\star}}{p_{\|}}\bdot\left({\bf K}^{\star}\btimes{\bf F}^{\star}\right) + \pd{{\bf K}^{\star}}{p_{\|}}\bdot\left({\bf F}^{\star}
  \btimes{\bf G}^{\star}\right)\right], 
\label{eq:Maxwell_two}
\end{eqnarray}
where $\leftturn$ denotes a cyclic permutation on the left side, we have used the cancellation identity \eqref{eq:Id_delta_B}, and $\partial{\bf K}^{\star}/\partial p_{\|} = (\partial\mathbb{P}_{\|}/\partial p_{\|})\bdot\nabla\btimes{\bf K}$.

\acknowledgments

This work was supported by the National Science Foundation grant No.~PHY-1805164.


\begin{thebibliography}{99}

\bibitem{Brizard_Tronci_2016} A.J.~Brizard and C.~Tronci, Phys.~Plasmas {\bf 23}, 062107 (2016).

\bibitem{Brizard_2021} A.J.~Brizard, {\it Hamiltonian structure of the guiding-center Vlasov-Maxwell equations}, in preparation (2021).

\bibitem{Morrison_2013} P.J. Morrison, Phys.~Plasmas {\bf 20}, 012104 (2013).

\bibitem{Brizard_2016} A.J.~Brizard, P.J.~Morrison, J.W.~Burby, L.~de Guillebon, and M.~Vittot, J.~Plasma Phys.~{\bf 82}, 905820608 (2016).


\end{thebibliography}
\end{document}